\documentclass[letterpaper,10pt,nofootinbib,aps,tightenlines,twocolumn]{revtex4}

\usepackage{amsmath,amsfonts,amssymb}
\usepackage{mathrsfs}
\usepackage{graphicx}
\usepackage[english]{babel} 
\usepackage{color}

\usepackage{appendix}

\usepackage{epstopdf}

\def\hatn{\mathbf{\hat n}}

\def\VEV#1{\left\langle #1 \right\rangle}
\def\Mpl{M_{\mathrm{ Pl}}}
\newcommand{\beq}{\begin{equation}}
\newcommand{\eeq}{\end{equation}}
\newcommand{\bga}{\begin{gathered}}
\newcommand{\ega}{\end{gathered}}
\newcommand{\beqa}{\begin{eqnarray}}
\newcommand{\eeqa}{\end{eqnarray}}

\begin{document}

\title{Cross-Correlation of Cosmological Birefringence with CMB
Temperature}
\author{Robert R.~Caldwell$^1$, Vera Gluscevic$^2$, and Marc
     Kamionkowski$^2$}
\affiliation{$^1$Department of Physics \& Astronomy, HB 6127
     Wilder Lab, Dartmouth College, Hanover, NH 03755, USA\\
     $^2$California Institute of Technology, Mail Code 350-17,
     Pasadena, CA 91125}

\date{\today}

\begin{abstract}
Theories for new particle and early-Universe physics abound
with pseudo-Nambu-Goldstone fields that arise when global
symmetries are spontaneously broken.  The coupling of these
fields to the Chern-Simons term of electromagnetism may give rise to
cosmological birefringence (CB), a frequency-independent
rotation of the linear polarization of photons as they propagate
over cosmological distances.  Inhomogeneities in the CB-inducing
field may yield a rotation angle that varies
across the sky.  Here we note that such a spatially-varying
birefringence may be correlated with the cosmic microwave
background (CMB) temperature.  We describe quintessence
scenarios where this cross-correlation exists and other scenarios
where the scalar field is simply a massless spectator field, in
which case the cross-correlation does not exist.  We discuss how the
cross-correlation between CB-rotation angle and CMB temperature may be
measured with CMB polarization.  This measurement may improve the sensitivity to the CB signal, and it can help discriminate between different models of CB.
\end{abstract}
\maketitle

\section{Introduction}
\label{sec:intro}

Much attention has focused recently on cosmological
birefringence (CB), a frequency-independent rotation of the linear
polarization of a photon that propagates over cosmological
distances \cite{Carroll:1989vb}.  The rotation may arise if the
pseudoscalar of electromagnetism ${F}^{\mu\nu}\widetilde{F}_{\mu\nu}$ is coupled to a
pseudo-Nambu-Goldstone field (PNGB field) that has variations on cosmological
distances or timescales.  This field may be identified with
the quintessence field \cite{quintessence,Caldwell:2009ix}
introduced to account for cosmic acceleration
\cite{Riess:1998cb,Perlmutter:1998np,Frieman:2008sn}. In fact,
the flatness required of the quintessence potential is naturally
accommodated if quintessence is a PNGB field
\cite{Carroll:1998zi, Frieman:1995pm}.
However, the scalar field  may have nothing to do with
quintessence---any PNGB field is
expected to have such a coupling \cite{Pospelov:2008gg}.  There
may also be dark-matter mechanisms for CB \cite{Gardner:2006za}.

Cosmological birefringence has been sought with polarized
cosmological radio sources
\cite{Carroll:1989vb,radio,Kamionkowski:2010ss}, but here we
focus on cosmic-microwave-background (CMB) probes of CB
\cite{Lue:1998mq}.  If the CB-rotation angle $\alpha$
is uniform across the sky, as may result from the homogeneous
evolution of quintessence, then there are parity-violating EB
and TB correlations between the CMB temperature (T) and the
curl-free (E) and curl (B) components of the
CMB polarization \cite{Lue:1998mq}.  Such a rotation has been
sought for several years \cite{earlysearches}, and the tightest
current limits on the rotation angle, $-1.41^\circ <
\alpha < 0.91^\circ$ (95\%\,C.L.), come from a combined
analysis of the WMAP \cite{Komatsu:2010fb}, BICEP
\cite{Chiang:2009xsa,Xia:2009ah}, and QUaD
\cite{Wu:2008qb,Brown:2009uy} experiments
\cite{Komatsu:2010fb}.  It is worth noting, in the context of
the current constraints, that the uniform rotation angle is
generally nonzero in quintessence models for CB, while the
massless-scalar-field models have no homogeneous time evolution
and thus predict no uniform rotation.

It has been pointed out that
\cite{Pospelov:2008gg,Gardner:2006za,Kamionkowski:2008fp},
more generally, the CB-rotation might be anisotropic, giving rise to $\alpha(\hatn)$ as a function of direction $\hatn$ in the sky.
Refs.~\cite{Kamionkowski:2008fp,Gluscevic:2009mm,Yadav:2009eb}
showed how measurement of the characteristic non-Gaussianities
in the CMB polarization induced by a spatially-varying CB
can be used to reconstruct $\alpha(\hatn)$ from the CMB.
The current best constraint to the root-variance of $\alpha$,
$\VEV{(\Delta \alpha)^2}^{1/2} \lesssim 4^\circ$, comes from
observations of active galactic nuclei \cite{Kamionkowski:2010ss}.

In this paper we explore the possibility that CB may be correlated with primordial density
perturbations and thus also with temperature fluctuations in the
CMB.  Such correlations are to be expected, for
example, if the CB-inducing field is a quintessence field with
adiabatic primordial perturbations seeded during inflation.
On the other hand, correlations between the CB angle and primordial perturbations may be
absent if, for example, the CB-inducing field is a
massless scalar~\cite{Pospelov:2008gg}.  

We first work out the predictions for spatially-varying
$\alpha(\hatn)$ for a massless-scalar-field model in
which there is no uniform rotation.  In this case, the
rotation-angle pattern $\alpha(\hatn)$ is completely
uncorrelated with the CMB-temperature pattern $T(\hatn)$, and
so we calculate only the CB-angle autocorrelation power spectrum
$C_L^{\alpha\alpha}$. We then move on to quintessence models in
which the $\alpha$T cross-correlation exists and calculate
this cross-correlation power spectrum $C_L^{\alpha T}$.  

We derive the minimum-variance estimators for the
$\alpha T$ cross-correlation and estimate the detectability
of CB-angle fluctuations and CB-angle--temperature correlations
with current and forthcoming CMB experiments. We find that the cross-correlation can help improve the sensitivity of experiments to a signal in some cases where the signal would
otherwise be only marginally detectable.  We show that experiments like
SPIDER \cite{SPIDER} and Planck \cite{Planck} may be able to
detect a cross-correlation if the CB signal is near its current
upper bounds, while the cross-correlation may be detectable with
a future experiment, like CMBPol/EPIC \cite{CMBPol}, even if the
CB power spectrum is several orders of magnitude smaller than
the current upper limit.

This paper is organized as follows:  In Section~\ref{sec:models}
we introduce the two anisotropic-CB scenarios and their
parameters and calculate the corresponding $\alpha\alpha$ and $\alpha T$ power spectra.
In Section~\ref{sec:constraints} we discuss how $\alpha$ can be
reconstructed from a CMB temperature/polarization map, and then
how the cross-correlations can be measured. Here, we present the expressions for the minimum-variance estimators of the auto-correlation and cross-correlation power spectra, and expressions for their variances.  We then evaluate
those variances for SPIDER, Planck, and CMBPol/EPIC and estimate
detectability thresholds for these three experiments. A summary and concluding remarks are presented in Section~\ref{sec:discussion}. 
Appendix \ref{sec:quintpert}
details the evolution of the scalar-field perturbations, and
Appendix \ref{sec:variance_alphaT_full_derivation} provides the
full expression for the variance of the $\alpha T$ cross-correlation.

\section{Scenarios for Anisotropic Rotation}
\label{sec:models}

We consider theories of a cosmic scalar $\phi(x^\mu)$ coupled to the Chern-Simons term of
electromagnetism via the Lagrangian
\begin{eqnarray}
\label{eq:FFdual}
     \mathscr{L} &=&
     -\frac{1}{2}(\partial_\mu\phi)( \partial^\mu\phi) -V(\phi) \cr
     &&-\frac{1}{4}F^{\mu\nu}F_{\mu\nu}- \frac{\beta \phi}{2M}F^{\mu\nu}\widetilde{F}_{\mu\nu},
\end{eqnarray}
where $\widetilde{F}_{\mu\nu} = \epsilon_{\mu\nu\rho\sigma}F^{\rho\sigma}/2$ is the dual of the electromagnetic tensor, $\epsilon_{\mu\nu\rho\sigma}$ is the Levi-Civita tensor (totally antisymmetric), and $M$ is a parameter with dimensions of mass.  If
$\phi$ is a PNGB field, then $M$ is the vacuum
expectation value for the broken global symmetry, and $\beta$ is
a coupling \cite{Carroll:1998zi, Frieman:1995pm}. Such a parity-violating term in the Lagrangian introduces a
modification of Maxwell's equations that results in different
dispersion relations for left- and right-circularly polarized
photons.  Consequently, linearly-polarized electromagnetic waves that propagate over cosmological distances undergo
CB, a frequency-independent rotation of
the plane of polarization by an angle $\alpha$, where \cite{Carroll:1989vb}
\begin{eqnarray}
     \alpha &=& \frac{\beta}{M}\int d\tau
     \left(\frac{\partial}{\partial\tau} -
     \hatn\cdot\vec\nabla\right)\phi \cr
    &=& \frac{\beta}{M} \Delta\phi,
\label{eqn:rotationangle}
\end{eqnarray}
where $\Delta\phi$ is the change in $\phi$ over the photon
trajectory, and $\tau$ is the conformal time. For the CMB, the
polarization rotation is determined by the change in $\phi$ since
recombination, when the CMB polarization pattern was largely
established. 

Allowing for spatial fluctuations $\delta\phi$
in the cosmic scalar field, the anisotropy in the
CB-rotation angle is then $\Delta\alpha(\hatn) =
({\beta}/{M})\delta\phi(\hatn)$, evaluated at recombination.

Below we consider two scenarios for the scalar field.
In the first, the scalar is massless, with no
homogeneous time evolution, while in the second, the scalar is
quintessence.  In both cases, CB-angle fluctuations arise
from scalar-field fluctuations at the surface of last scatter (LSS).

\subsection{Massless scalar field}

In the first scenario, we suppose that the $\phi$ field is
simply a massless scalar with a potential that vanishes, $V=0$. In this case, the value of the field is completely uncorrelated with primordial density perturbations\footnote{We imagine that some mechanism has nullified the quantum-gravity effects that generically break
global symmetries \cite{Kamionkowski:1992mf}.}
\cite{Pospelov:2008gg}.
If $\phi$ is effectively massless during
inflation there will be a scale-invariant power spectrum of
perturbations to $\phi$, $P_{\delta\phi}(k) = H_I^2/2k^3$, with an amplitude fixed by the Hubble parameter $H_I$ evaluated during inflation\footnote{It is also imaginable that a white-noise spectrum of $\phi$ fluctuations is imprinted by some
post-inflation phase transition, but we will not consider that
scenario here.}. If we split the field into a smooth background component and a perturbation on top of it,  $\phi(\tau,\vec{x}) = \phi_0(\tau) + \delta\phi(\tau,\vec{x})$, the evolution of the homogeneous component is given by the following equation of motion
\begin{equation}
\label{eqn:phi0}
     \ddot\phi_0 + 2 {\cal H}\dot\phi_0 + a^2 V'=0,
\end{equation}
where ${\cal H}= \dot a/a$, $a$ is the scale factor, and dots denote derivatives with respect to conformal time. 
For a vanishing potential, this equation has only a decaying and a constant solution; thus, the value of the field is fixed in time in each causally disconnected region of the early Universe. This precludes the scalar-field perturbations from having any correlation with perturbations in the matter/radiation
density.  This is manifest in the absence of any source term in
the perturbed equation of motion for the scalar field (compare to the Fourier transform of the full equation, Eq.~(\ref{eqn:deltaphitwo}), after taking $(d\phi/d\tau)=0$, and $V=0$), 
\begin{equation}
\label{eqn:deltaphitwo_nosource}
     \delta\ddot\phi + 2 {\cal H}\delta\dot\phi - k^2\delta\phi =  0.
\end{equation}
A solution to Eq.~(\ref{eqn:deltaphitwo_nosource}) is a transfer function $T_k(\tau) \propto j_1(k\tau)/(k\tau)$, which describes the conformal-time evolution of a given Fourier mode of wavenumber $k$ during matter domination.

The angular power spectrum
$C_L^{\alpha\alpha}$ for the rotation angle is then
\begin{eqnarray}
     C^{\alpha\alpha}_{L} &=& 4\pi
     \left( \frac{\beta}{M} \right)^2 \int \frac{k^2 \,
     dk}{2\pi^2} P_\phi(k) \left[ j_L(k\Delta\tau) 
     T_k(\tau_{lss}) \right]^2 \cr 
     &=& \frac{1}{\pi}
     \left(\frac{\beta H_I}{M}\right)^2  \int
     \frac{dk}{k} \left[j_L(k\Delta\tau) T_k(\tau_{lss})\right]^2.
\label{eqn:alphapower}
\end{eqnarray}
Here $\Delta\tau$ is the conformal-time difference between last
scattering and today, and $\tau_{lss}$ is the conformal time at
the LSS.  For large angular scales, $L\lesssim 100$, the
transfer function evaluates to $T_k(\tau_{lss})\simeq 1$, in which
case
\begin{equation}
     C^{\alpha\alpha}_L \simeq  \frac{(\beta H_I/M)^2 }{2\pi
     L(L+1)}, \qquad \mathrm{for}\, L \lesssim 100.
\label{eqn:Clapprox}
\end{equation}
The left-hand panel of Fig.~\ref{fig:s_to_n_100_toy} shows the result of a numerical
calculation of $C_L^{\alpha\alpha}$ for this scale-invariant
power spectrum.

The mean-square rotation amplitude measured by a probe with
angular resolution of $\sim1^\circ$ is
\begin{eqnarray}
     \VEV{ (\Delta\alpha)^2 } &=&
     \sum_{L=2}^{\infty}  \frac{2L+1}{4\pi}
     C_{L}^{\alpha\alpha} \left[W_L(\theta)\right]^2   \nonumber \\
     &\simeq &
     332 \left(\frac{\beta H_I}{M}\right)^2 \, {\rm deg}^2.
\label{eqn:rms}
\end{eqnarray}
Here, $W_L(\theta) \equiv \exp\left[-L^2 \theta^2/(16\ln 2)
\right]$ is a Gaussian window function of full-width
half-maximum $\theta$ (in radians).  The best current constraint to the
variance of the rotation
angle, $\VEV{(\Delta\alpha)^2} \lesssim (4^\circ)^2$, from AGN
data \cite{Kamionkowski:2010ss}, places a
bound $\beta H_I/M \lesssim 0.2$ to the combination of
parameters that control the rotation-angle amplitude in this
scenario.  We may therefore write the rotation-angle power
spectrum as
\begin{eqnarray}
     C^{\alpha\alpha}_{L} &=& 0.015\,\alpha_4^2
     \int \frac{dk}{k} \left[ j_L(k\Delta\tau) 
     T_k(\tau_{ls}) \right]^2 \cr 
     &\simeq& \frac{7.7\times 10^{-3} \alpha_4^2}{
     L(L+1)}, \qquad \mathrm{for}\, L \lesssim 100.
\label{eqn:alphasimple}
\end{eqnarray}
Here, $\alpha_4\equiv
\VEV{(\Delta\alpha)^2}^{1/2}/4^\circ$ parametrizes the
amplitude of the rotation-angle power spectrum in units of the
maximum value currently allowed.  

The CMB temperature power spectrum is given by
\begin{equation}
     C_L^{TT} = \frac{2}{\pi} \int\, k^2\, dk\,
     \left[\Delta_{T,L}(k) \right]^2 P_\Psi(k),
\label{eqn:Tpowerspectrum}
\end{equation}
where $P_\Psi(k)$ is the primordial power spectrum for the gravitational potential, and $\Delta_{T,L}(k)$ is the transfer function that
quantifies the contribution of a density mode of wavenumber $k$
to $C_L^{TT}$, and may be obtained from numerical Boltzmann
codes \cite{cmbfast}.  

As discussed above, scalar-field fluctuations are not sourced by
the gravitational potentials for this $V=0$ model; similarly, 
energy-density fluctuations in the scalar field have only second-order corrections due to $\delta\phi$, and so their effect on
gravitational potentials is also small.  In this case, the $\alpha$T
cross-correlation power spectrum vanishes, $C_L^{\alpha T}=0$.

\subsection{Quintessence}

In the second scenario, we suppose that $\phi$ is a quintessence
field with a nonzero potential and homogeneous component that
undergoes time evolution.  In this case, gravitational-potential
perturbations directly source (and are also sourced by)
scalar-field fluctuations, see Eq.~(\ref{eqn:deltaphitwo}).  A cross-correlation between CB-angle
and CMB-temperature fluctuations is therefore inevitable,
although its amplitude and detailed features depend on the
specific potential $V$.

Since every CMB photon that comes from a given direction $\hatn$ last
scattered at the spacetime point in the direction $\hatn$, when
the Universe had some fixed temperature, the CB-rotation angle
$\alpha(\hatn)$ is determined by the value of
$\phi$ at that point of spacetime.  In other words, the CB-angle anisotropies
are determined by the scalar-field perturbations on surfaces of
constant CMB temperature, or equivalently, on surfaces of
constant synchronous-gauge time.

We suppose that the initial value of $\phi$ is
set by some post-inflationary physics so that the
primordial perturbation to $\phi$ is adiabatic.  In this case,
the synchronous-gauge scalar-field perturbation
$(\delta\phi)_{\mathrm{syn}}$ is initially zero.  However, the
scalar-field perturbation is sourced by the gravitational
potentials, as described by Eqs.~(\ref{eqn:deltaphione}) or
(\ref{eqn:deltaphitwo}).  The synchronous-gauge scalar-field
perturbation at the LSS is then
approximately (see Appendix \ref{sec:quintpert}),
\begin{equation}
     (\delta \phi)_{\mathrm{syn,lss}} = -\frac{2}{9}\left(\frac{3\Omega_\phi
     (1+w_\phi)}{8\pi}\right)^{1/2} \Mpl \Psi,
\label{eqn:deltaphi}
\end{equation}
where the equation-of-state parameter $w_\phi$ and the energy-density parameter $\Omega_\phi$ are evaluated at recombination. The primordial power spectrum for the gravitational
potential, for large scales (small $k$) is given by
\begin{equation}
     P_\Psi = \frac{9}{25} \frac{2 \pi^2}{k^3}\Delta_{\cal R}^2,
\label{eqn:Ppsi_primordial}
\end{equation}
where we have taken a scalar spectral index to be $n_s=1$ for
simplicity, and the curvature-perturbation amplitude is
$\Delta_{\cal R}^2(k_0) = 2.43(\pm0.11)\times 10^{-9}$
\cite{Larson:2010gs}. To evolve the power spectrum from primordial to the LSS,we need to multiply it by transfer functions, which are a suppression factor for small scales (large $k$'s).  The angular power spectrum for the CB-rotation angle in the quintessence model is then
\begin{eqnarray}
     C^{\alpha\alpha}_{L} &=&  \frac{2}{27} \Omega_\phi (1+w_\phi) \left(
     \frac{\beta \Mpl}{M} \right)^2 \cr
     & & \times  \int \frac{k^2\, dk}{2\pi^2}
     P_\Psi(k) [j_L(k\Delta\tau) T_k(\tau_{lss})]^2.
\label{eqn:QPS}
\end{eqnarray}
For large scales, $L\lesssim 100$, we can approximate
$T_k(\tau_{lss})\simeq 1$, in
which case we can again write the CB-rotation--angle power spectrum
as in Eq.~(\ref{eqn:alphasimple}), but now with
\begin{equation}
     \alpha_4 \simeq  6.7\times10^{-5}\, \sqrt{  \Omega_\phi(1+w_\phi)}
     (\beta \Mpl/M).
\label{eqn:alphaQ}
\end{equation}
In other words, the $\alpha\alpha$ power spectrum for the
quintessence scenario will be similar to that for the
massless-scalar-field scenario in the small-$L$ limit where $T_k(\tau_{lss})$ can be approximated as a constant.

\begin{figure}
\includegraphics[height=8.5cm,keepaspectratio=true]{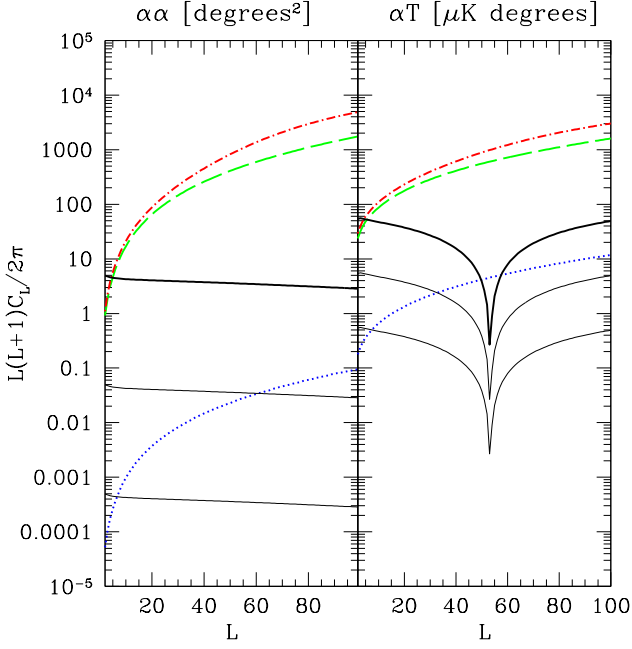}
\caption{Shown are the power spectra
     for the cosmological-birefringence rotation
     angle $C_L^{\alpha\alpha}$ and its cross-correlation with the CMB
     temperature $C_L^{\alpha T}$ (logarithm of the absolute value), for a generic quintessence model in which
     the CB-angle fluctuations are due to scalar-field fluctuations
     at the LSS.  The black solid curves are
     the theoretical prediction for
     (from top to bottom) $\alpha_4=1$, 0.1, and 0.01, where
     $\alpha_4$ is the fluctuation amplitude for the CB angle in
     units of the maximum currently allowed amplitude
     \protect\cite{Kamionkowski:2010ss}.  We also
     show the noise power spectra anticipated for SPIDER (red,
     dot-dashed), Planck (green, dashed), and
     CMBPol (blue, dotted).\label{fig:s_to_n_100_toy}}
\end{figure}

However, in the quintessence model, there will also be a
cross-correlation with the CMB temperature, since the CMB
temperature is determined largely by the potential $\Psi$ at the
LSS.  From Eqs.~(\ref{eqn:Tpowerspectrum})
and (\ref{eqn:QPS}), we get
\begin{eqnarray}
     C_L^{\alpha T} &=& -\frac{4 \pi}{3} \sqrt{ \frac{\Omega_\phi
     (1+w_\phi)}{6\pi}} 
     \frac{\beta \Mpl}{M} \cr
     & & \times \int \frac{k^2 \, dk}{2\pi^2}
     P_\Psi(k) \Delta_{T,L}(k) j_L(k\Delta\tau)
     T_k(\tau_{ls}). \nonumber \\
\label{eqn:cross}
\end{eqnarray}
The absolute value of this cross-correlation is also
shown in Fig.~\ref{fig:s_to_n_100_toy}.  The passage through zero
at $L\sim 50$ arises because of the relative contributions of
the monopole and dipole contributions of the photon distribution
function to $\Delta_{T,L}(k)$.

\section{Prospects for Detection}\label{sec:constraints}

In this Section, we first review the procedure presented in
Refs.~\cite{Kamionkowski:2008fp,Gluscevic:2009mm,Yadav:2009eb}
for constructing the minimum-variance estimator for the CB-rotation angle from
a CMB temperature/polarization map.  We then extend this work to show how the
cross-correlation with the temperature can be reconstructed.  
At the end, we evaluate the detectability of the CB-rotation
with both the auto-correlation and the cross-correlation for the
CB scenarios discussed in Section~\ref{sec:models}.

\subsection{Measuring the rotation angle}
\label{var_alphaT_derivation}

Refs.~\cite{Kamionkowski:2008fp,Gluscevic:2009mm}
show how the CB-rotation--angle spherical-harmonic coefficients
$\alpha_{LM}$ can be reconstructed from a full-sky CMB
temperature/polarization map.  While these coefficients can be
obtained from EE, TE, TB, and EB cross-correlations, the best
sensitivity will ultimately come from the EB
cross-correlation.  We therefore restrict our attention to
reconstruction of $\alpha(\hatn)$ from the EB power spectra.

To begin, the E/B spherical-harmonic coefficients,
$E_{lm}^{\mathrm{map}}$ and $B_{lm}^{\mathrm{map}}$, are
constructed from the full-sky map of the Stokes parameters,
$Q(\hatn)$ and $U(\hatn)$, in the usual way
\cite{Kamionkowski:1996ks,Zaldarriaga:1996xe}.  Following
Refs.~\cite{Kamionkowski:2008fp,Gluscevic:2009mm}, the
minimum-variance estimator for the rotation-angle
spherical-harmonic coefficient is
\begin{eqnarray}
     \widehat{\alpha}_{LM} & =& C_L^{\alpha\alpha,\mathrm{noise}}\sum_{mm',l' \geq
     l} \xi^{LM}_{lml'm'}[V^{L}_{ll'} E^{\text{map}}_{l'm'}
     B^{\text{map}}_{lm} \nonumber \\
     & & \qquad\qquad\qquad + V^{L}_{l'l}E^{\text{map}}_{lm}B^{\text{map}}_{l'm'}],
\label{eqn:alpha_estimator}
\end{eqnarray}
where
\begin{equation}
     V^{L}_{ll'}\equiv \frac{F^{L, BE}_{ll'}}{(1+\delta
     _{ll'})C^{BB\text{, map}}_l C^{EE \text{, map}}_{l'}},
\end{equation}
\begin{equation}
	F^{L, BE}_{ll'}\equiv 2C^{EE}_{l'} 
\left( {\begin{array}{*{20}c}
           l & L & l'  \\
              2 & 0 & -2  \\
              \end{array}} \right) W_lW_{l'}, \quad
	F^{L, EB}_{ll'}\equiv F^{L, BE}_{l'l} ,
\end{equation}
and
\begin{eqnarray}
        \xi^{LM}_{lml'm'} &\equiv&
        (-1)^m\sqrt{\frac{(2l+1)(2l'+1)(2L+1)}{4\pi}} \nonumber
        \\
        & & \times \left(
        {\begin{array}{*{20}c}   l & L & l'  \\
   -m & M & m' \\
\end{array}} \right).
\end{eqnarray}
Here, the objects in parentheses are Wigner-3j symbols, and $W_l$ is the window function defined in Section \ref{sec:models}. $C_l^{EE,\mathrm{map}}$ and $C_l^{BB,\mathrm{map}}$ are,
respectively, power spectra for the E and B modes from the map
(including instrumental noise); i.e.,
\begin{equation}
     C^{XX'\text{, map}}_l\equiv C^{XX'}_l|W_l|^2 +
     C^{XX'\text{, noise}}_l,
\label{Cobs}
\end{equation}
where $XX'\in \{TT, EE, BB, ET, EB, TB\}$.
The noise power spectra are
\begin{equation}
\begin{gathered}
  C^{TT\text{, noise}}_l \equiv \frac{4\pi f^0_\text{sky}(\text{NET})^2}{t_{\text{obs}}},\\
  C^{EE\text{, noise}}_l=C^{BB\text{, noise}}_l\equiv 2C^{TT\text{, noise}}_l,\\ 
  C^{EB\text{, noise}}_l=C^{TB\text{, noise}}_l\equiv 0,
\end{gathered} 
\label{Cnoise}
\end{equation}
where $t_{\text{obs}}$ is the total observation time, $f^0_\text{sky}$ is the fraction of the sky surveyed (taken to be different from $1$ only for SPIDER, where $f^0_\text{sky}=0.5$), and NET is the noise-equivalent temperature. We assume no
cross-correlation between the noises in polarization and
temperature and apply the null assumption (no B modes in the
signal), so there are no TB and EB correlations.  The power
spectrum $C^{BB\text{, map}}_l$ thus contains only the contribution
from instrumental noise.

Under the null hypothesis of no rotation, the expectation value
of the estimator in Eq.~(\ref{eqn:alpha_estimator}) is zero, and
its variance is the $\alpha\alpha$ noise power spectrum as given in Ref.~\cite{Gluscevic:2009mm},
\begin{eqnarray}
    C_L^{\alpha\alpha,\mathrm{noise}} &\equiv& \VEV{ |\widehat
    \alpha_{LM}|^2}  \cr
    &=& \left[ \sum_{ll'} 
     \frac{(2l+1)(2l'+1)(F^{L,BE}_{ll'})^2}{ 4 \pi C^{BB\text{,
     map}}_l C^{EE \text{, map}}_{l'}} \right]^{-1}.
\label{eqn:alpha_variance}
\end{eqnarray}

If the polarization pattern at the LSS is a realization of a statistically isotropic field,
then there are $2L+1$ statistically independent $M$ modes for
each $L$ in $\widehat \alpha_{LM}$. In this case, each $M$ mode provides an independent estimator of the rotation power spectrum, $C_L^{\alpha\alpha}$. The minimum-variance estimator is then
\begin{equation}
     \widehat C^{\alpha \alpha}_L=\frac{1}{2L+1} \sum_{M=-L}^{L
     }|\widehat{\alpha}_{LM} |^2.
\label{eqn:Calphaalpha_estimator}
\end{equation}
Each $\widehat{\alpha}_{LM}$ is a sum of products of Gaussian
random variables and is thus not a Gaussian random
variable.  However, if the number of terms in the sum is large,
the central-limit theorem holds, and $\widehat{\alpha}_{LM}$ can
be approximated as Gaussian. In this case, the
expression for the variance of $\widehat C^{\alpha \alpha}_L$
takes on the usual form,
\begin{equation}
     \left( \Delta \widehat C_L^{\alpha\alpha} \right)^2 \simeq
     \frac{2}{f_\text{sky}(2L+1)} \left(C_L^{\alpha\alpha,\mathrm{noise}}
     \right)^2,
\label{eqn:Calphaalpha_variance}
\end{equation}
where $f_\text{sky}$ is the sky-cut used in the analysis, taken to be $0.8$ for Planck and CMBPol and $0.5$ for SPIDER.

\subsection{Measurement of the rotation-temperature cross-correlation}

In analogy with the derivation in
Ref.~\cite{Kamionkowski:1996ks} of the estimator for $C_l^{TE}$,
the estimator for $C_L^{\alpha T}$ is
\begin{equation}
     \widehat C^{\alpha
     T}_L=\frac{1}{2L+1}\sum_{M=-L}^{L}\widehat{\alpha}_{LM}
     (T^{\mathrm{map}}_{LM})^* W_L^{-1},
\end{equation}
where $T^{\mathrm{map}}_{LM}$ is the temperature
spherical-harmonic coefficient obtained from the map.  
Under the null hypothesis, $\widehat T_{LM}$ has no correlation with any
$B_{LM}$s, and it is correlated with $E_{LM}$ with the same $L$
and $M$, but uncorrelated with any other $E_{LM}$.  The
estimator $\widehat\alpha_{LM}$ depends on a large number of
$E_{lm}$'s but does not include $\{lm\}=\{LM\}$.  There is
therefore no correlation (under the null hypothesis) of
$\widehat\alpha_{LM}$ and $\widehat T_{LM}$; i.e., 
there is no noise contribution to $C_L^{\alpha T}$.
Again, if $\widehat\alpha_{LM}$ is approximately Gaussian, then
the variance with which $C_L^{\alpha T}$ can be measured is
approximately that obtained assuming $\widehat\alpha_{LM}$ is
Gaussian. To check the validity of this assumption for the purpose of calculating the sensitivity of future CMB experiments to the CB signal (see Section \ref{alphaT_vs_alphaalpha}), we evaluate the full expression for this variance (without assuming Gaussianity of $\widehat\alpha_{LM}$, see Eq.~(\ref{sigma_alphaT_full_expression})) and confirm that the numerical results agree up to a level of a few percent. Thus, for simplicity, and without any loss in accuracy, we can invoke analogy with the variance of $C_l^{TE}$ (see,
e.g., Ref.~\cite{Kamionkowski:1996ks}) to get
\begin{equation}
     \left( \Delta C_L^{\alpha T} \right)^2 \simeq
     \frac{1}{f_\text{sky}(2L+1)} C_L^{\alpha\alpha,\mathrm{noise}}
     C_L^{TT,\mathrm{map}} W_L^{-2}.
\label{eqn:CalphaT_variance}
\end{equation}

\subsection{Sensitivity to Detection: $\alpha T$ vs. $\alpha\alpha$}
\label{alphaT_vs_alphaalpha} 

We now return to our two models for CB which predict that the
rotation $\alpha$ is a realization of a random field with the
power spectra $C_L^{\alpha\alpha}$ and $C_L^{\alpha T}$ presented in
Fig.~\ref{fig:s_to_n_100_toy}.  Our aim here is to
evaluate the smallest signal amplitude
detectable by measurement of the rotation alone, as well as the
smallest amplitude detectable by measurement of the
rotation-temperature cross-correlation.

We write the power spectra as $C_L^{\alpha
\alpha}\equiv \alpha_4^2
C^{\alpha\alpha\text{,fiducial}}_L$, and $C^{\alpha T}_L\equiv
\alpha_4 C^{\alpha T\text{,fiducial}}_L$,
where the fiducial model ($\alpha_4=1$) is the
quintessence model in Fig.~\ref{fig:s_to_n_100_toy}
with the largest amplitude allowed by current rotation-angle constraints.
The inverse-variance with which the amplitude $\alpha_4^2$ of
the $\alpha\alpha$ power spectrum can be obtained
from the rotation-angle auto-correlation is
\cite{Jungman:1995bz}
\begin{eqnarray}
   \frac{1}{\left [\Delta(\alpha_4^2) \right]^2} &=&
     \sum_L \left(\frac{\partial C^{\alpha \alpha}_L}{\partial (\alpha_4^2)}
     \right)^2 \frac{1}{\left( \Delta \widehat C_L^{\alpha\alpha} \right)^2}\cr
	  &=&\sum_L
          \left(\frac{C^{\alpha\alpha\text{,fiducial}}}{\Delta
          \widehat C_L^{\alpha\alpha}}
     \right)^2.
\label{eqn:sigma_Aalphaalpha}
\end{eqnarray}
Similarly, the inverse-variance with which the amplitude $\alpha_4$ of the $\alpha T$ power spectrum can be obtained from the cross-correlation of the rotation with the
temperature is
\begin{eqnarray}
     \frac{1}{\left(\Delta\alpha_4 \right)^2} &=&
     \sum_L \left(\frac{\partial C^{\alpha T}_L}{\partial \alpha_4}
     \right)^2 \frac{1}{\left( \Delta \widehat C_L^{\alpha T} \right)^2} \cr\cr
	 &=&\sum_L \left(\frac{C^{\alpha
         T\text{,fiducial}}}{\Delta \widehat C_L^{\alpha T}}
     \right)^2.
\label{eqn:sigma_AalphaT}
\end{eqnarray}
From these relations, we can estimate the signal-to-noise ratio for
measurement of $\alpha_4^2$ from the $\alpha\alpha$
autocorrelation to be $(S/N)_{\alpha\alpha} =
\alpha_4^2/\left [\Delta(\alpha_4^2) \right]$ and a
signal-to-noise for measurement of $\alpha_4$ from the
$\alpha$T cross-correlation to be $(S/N)_{\alpha\mathrm{T}} =
\alpha_4/ (\Delta\alpha_4)$.  We evaluate these expressions for our fiducial model ($\alpha_4=1$), for different instrumental parameters in Section \ref{var_numerics}. The smallest
$\alpha_4$ detectable at the $2\sigma$ level from the
cross-correlation and auto-correlation are then $2
\Delta\alpha_4$ and $\left[2 \Delta(\alpha_4^2)
\right]^{1/2}$, respectively.

\begin{table}[htbp]
\begin{center}
    \begin{tabular}{ | c | c | c | c | c | c | p{10cm} |}
    \hline
    Instrument & $\theta$ & NET &
    $t_{\text{obs}}$ & $(S/N)_{\alpha\alpha}$ & $(S/N)_{\alpha
    T}$  \\ \hline
    SPIDER & 60 & 3.1 & 0.016 & 9 & 7 \\ \hline
    Planck & 7.1 & 62 & 1.2 & 11 & 9 \\ \hline
    CMBPol/EPIC & 5 & 2.8 & 4 & $2\times 10^5$ &
    $1200$ \\ \hline
    \end{tabular}
\caption{Instrumental parameters from Refs.~\cite{SPIDER,
     Planck, CMBPol} for the three experiments considered in
     this work: beamwidth $\theta$ (in arcminutes),
     noise-equivalent temperature (NET) (in $\mu$K~sec$^{1/2}$),
     and observation time $t_{\text{obs}}$ (in years). The last
     two columns list signal-to-noise
     ratios (S/N) for the CB-angle
     auto-correlation and its cross-correlation with the CMB
     temperature, for our fiducial quintessence model ($\alpha_4=1$) shown in
     Fig.~\protect\ref{fig:s_to_n_100_toy}.  Note that
     the signal-to-noise scales with the signal amplitude
     $\alpha_4$ as $(S/N)_{\alpha\alpha} \propto \alpha_4^2$ and
     $(S/N)_{\alpha\mathrm{T}} \propto \alpha_4$.}\label{instruments}
\end{center}
\end{table}

\subsection{Numerical Results}\label{var_numerics}

We now present numerical results for the $\alpha\alpha$ and
$\alpha T$ noise power spectra and evaluate the largest possible
signal-to-noise and the smallest
detectable amplitude $\alpha_4$ for three CMB
polarization experiments:
(i) SPIDER's 150 GHz channel \cite{SPIDER}, (ii) Planck's 143 GHz
channel \cite{Planck}, and (iii) CMBPol's (EPIC-2m) 150 GHz
channel \cite{CMBPol}.  We obtain the CMB
temperature-polarization power spectra from CMBFAST
\cite{cmbfast} using WMAP-7 cosmological parameters
\cite{Komatsu:2010fb}. The instrumental parameters we use are
listed in Table \ref{instruments}.
Fig.~\ref{fig:s_to_n_100_toy} shows the noise power spectra\footnote{Note that there is a difference in normalization between the \textit{noise} and the \textit{variance}: $C_L^{XX',\mathrm{noise}}\equiv\sqrt{(2L+1)/2}\Delta \widehat C^{XX'}_L$, where $XX'=\{\alpha\alpha , \alpha T\}$. It is customary to plot the noise power spectra, even though the variance enters the expressions for signal-to-noise.} 
$C_L^{\alpha\alpha,\text{noise}}$ and $C_L^{\alpha 
T,\text{noise}}$.  For $C^{\alpha T}_L$, strictly speaking,
there is no instrumental-noise contribution, only the
effective noise, arising from
cosmic variance.  Table
\ref{instruments} lists signal-to-noise ratios, assuming
$\alpha_4=1$, for the auto- and cross-correlations, for these three experiments.  We find that SPIDER and Planck may already have
the sensitivity to detect not only the signal, but also its
cross-correlation with with the temperature, in the best-case
scenario of $\alpha_4\simeq1$, where the signal is just below the current detection limit\footnote{Here we have 
assumed that the errors to the rotation-angle
estimators are approximately Gaussian. However, if the signal is
just detectable (for example, for experiments like SPIDER if
$\alpha_4\simeq 1$), then this assumption may break down, and if
so, the precise quantitative forecasts for the signal-to-noise
may differ slightly.}.  In both cases, the sensitivity to
the signal may be improved if the auto- and cross-correlations
are measured in tandem.  CMBPol should have sensitivity to a
signal as small as $\alpha_4\sim 10^{-5}$, and a
detection of the cross-correlation of very high signal-to-noise
may be obtained with CMBPol if $\alpha_4\simeq1$.

\section{Summary and Discussion}
\label{sec:discussion}

If a quintessence field gives rise to cosmological
birefringence, then a correlation between CB--rotation-angle fluctuations
and CMB-temperature fluctuations is
inevitable.  We calculated that cross-correlation assuming the initial quintessence
perturbations are adiabatic.  We also discussed, by way of
contrast, a scenario in which the CB-inducing field is just a
massless scalar field that has no correlation with primordial
perturbations.

We derived the minimum-variance estimator for
the $\alpha T$ power spectrum that can be obtained from a CMB temperature-polarization map.  We find that
measurement of this cross-correlation may improve sensitivity to
the CB signal in some cases where the signal would otherwise be
only marginally detectable.  We further show that a high
signal-to-noise measurement of this cross-correlation is
conceivable with forthcoming and future CMB experiments if the
rotation-angle power-spectrum amplitude is near its current
upper limit.  Measurement of this cross-correlation may thus
provide another empirical handle with which to discover new physics indicated by cosmological birefringence.

We have restricted our attention to the EB estimator for the
rotation angle, as it is expected to provide the best
sensitivity.  However, there may be some improvement, though
probably small, with the inclusion of the TE, TB, and EE
estimators for the rotation.  We leave this calculation for
future work.  Likewise, we have left more careful investigation
of the impacts of partial-sky analysis, foregrounds, uneven
noise, as well as the details of distinction of these
signals from weak lensing, for future work.

We have refrained from discussing details of the quintessence
model here, as the angular dependence of the CB power spectra at
superhorizon scales at the time of recombination,
$L\lesssim100$, is insensitive to these details.  The dependence
of the amplitudes of the $\alpha\alpha$ and $\alpha$T power
spectra is given in terms of the quintessence parameters
$\Omega_\phi$ and $w$ at the LSS by
Eqs.~(\ref{eqn:QPS}) and (\ref{eqn:cross}).  However, if the quintessence field
couples to the pseudoscalar of electromagnetism, it is natural
to expect it to be a pseudo-Nambu-Goldstone field, and if so,
then its potential should be $V(\phi) \propto
[1-\cos(\phi/f)]$.  In this case, the quintessence field $\phi$
is frozen at early times leading to spatial variations in
$\alpha$ that are unobservably small.  In this case, though,
additional fluctuations in $\alpha$ may be produced during the
epoch of reionization \cite{inpreparation}.

For the massless scalar field, the uniform CB-rotation angle is
expected to be zero, and so a search for the fluctuations
is essential to detect the signal.  For quintessence,
however, the uniform rotation is expected to be nonzero and
generically quite a bit larger than the fluctuations, which, given the current best constraint may imply a relatively small amplitude of the fluctuations power spectrum.  However,
CMBPol may be sensitive to a fluctuation amplitude as small as
$\sim10^{-5}$ of the current upper limit to the uniform
rotation, which, if detected, would help distinguish between different
CB scenarios. Moreover, the fluctuation amplitude in the quintessence
scenario could be larger than a measured uniform-rotation angle.
This could occur if, for example, the uniform-rotation angle
(which can only be recovered mod $\pi$) happens to be close to
an integer multiple of $\pi$.  It will be interesting, with
forthcoming precise CMB maps, to address these questions
empirically rather than through theoretical speculation.

\begin{acknowledgments}
We thank C. Hirata for useful discussions. MK thanks the support of the Miller Institute for Basic Research in Science and the hospitality of the Department of Physics at
the University of California, Berkeley, where part of this work
was completed.  This work was supported in part by NSF
AST-0349213 at Dartmouth and by DoE DE-FG03-92-ER40701, NASA
NNX10AD04G, and the Gordon and Betty Moore Foundation at Caltech.
\end{acknowledgments}

\appendix

    
\section{Quintessence Perturbations}
\label{sec:quintpert} 

As discussed in the paper, the CB-angle fluctuation is determined by
the synchronous-gauge scalar-field fluctuation
$(\delta\phi)_{\mathrm{syn}}$ at the LSS.
To obtain this fluctuation, we start from adiabatic initial conditions, and then evolve the scalar-field--perturbation equation of motion forward in time, from the early radiation-dominated epoch to the LSS.
The equation of motion is
\begin{equation}
\label{eqn:deltaphione}
     \delta\ddot\phi + 2 {\cal H}\delta\dot\phi + a^2 
     V''\delta\phi - \nabla^2\delta\phi =-\frac{1}{2}\dot h
     \dot\phi,
\end{equation}
in the synchronous gauge, and
\begin{equation}
\label{eqn:deltaphitwo}
     \delta\ddot\phi + 2 {\cal H}\delta\dot\phi + a^2
     V''\delta\phi - \nabla^2\delta\phi 
     =  \dot\phi (3\dot\Phi+ \dot\Psi) - 2 a^2 V' \Psi,
\end{equation}
in the conformal-Newtonian/longitudinal gauge. (See
Ref.~\cite{Ma:1995ey} for definitions of the metric variables $\Phi$, $\Psi$, $\eta$, and $h$.) Adiabatic initial conditions require that the perturbations of the scalar field vanish at early times. However, the subsequent evolution of the scalar field is not adiabatic, meaning that $(\delta\phi)_{\mathrm{syn, lss}}$ does not necessarily vanish at the LSS, even though all the matter and radiation perturbations do. All of our numerical integrations that give the power spectra presented in the Figure are done in the synchronous gauge, using a modified version of CMBFast \cite{cmbfast}.  

For the numerical work, we assume a
quintessence potential of the PNGB form
$V(\phi) = m^4 (1-\cos\phi/f)$, as expected if $\phi$ is an
axion-like field.  We take an initial value
$\phi$, $m$, and $f$ so that
$\Omega_\phi=0.7$ today and the density-weighted average equation-of-state parameter
$\langle w \rangle \simeq-0.95$, which gives
$\Delta\phi=0.045\,\Mpl$ for the change in the scalar field
between decoupling and today.  However, the numerical results
presented in the Figure will be similar for any quintessence
potential that has $w_\phi\to-1$ at early times.

The numerical results can be largely reproduced with the analytic approximation for
$(\delta\phi)_{\mathrm{syn, lss}}$, given in Eq.~(\ref{eqn:deltaphi}),
which we now derive.  We now work in the conformal-Newtonian/longitudinal gauge and we make the
approximation that decoupling takes place well into matter
domination; we assume that most of the growth in perturbations happens during this epoch. For $w_\phi\to-1$, the $V''$ term in Eq.~(\ref{eqn:deltaphitwo}) is negligible. Additionally, in the superhorizon limit, valid for multipoles $L\lesssim100$, we can neglect the spatial-gradient term.  The
simplified equation of motion is then,
\begin{equation}
\label{eqn:deltaphapprox}
     \delta\ddot\phi + 2 {\cal H}\delta\dot\phi 
     \simeq - 2 a^2 V' \Psi.
\end{equation}
Aside from the homogeneous solutions that are either constant or decaying, it also has an inhomogeneous solution that grows as
\begin{equation}
     (\delta\phi)_{\mathrm{con}} \simeq - a^2 \tau^2 V' \Psi/27,
\end{equation}
during matter domination.  The potential derivative $V'$ can be
expressed, using the quintessence slow-roll approximation, from 
\begin{equation}
a^2 V'\simeq -3 {\cal H} \dot\phi.
\label{eqn:slowrollVp}
\end{equation}
Also, 
\begin{equation}
\dot\phi^2 = a^2 \rho_\phi(1+w_\phi),
\end{equation} 
with $\rho_\phi = \Omega_\phi \rho_c$ and $\rho_c = 3H^2M_{Pl}^2/(8\pi)$.  We then find
\begin{equation}
     (\delta\phi)_{\mathrm{con}} = \frac{4}{9} \left[
     \frac{3}{8\pi} \Omega_\phi (1+w_\phi) \right]^{1/2} M_{Pl}
     \Psi.
\end{equation}
The result in Eq.~(\ref{eqn:deltaphi}) is then obtained by going back to the synchronous gauge, using the gauge-transformation equations \cite{Ma:1995ey}
\begin{eqnarray}
     (\delta\phi)_{\mathrm{syn}}  &=&
     (\delta\phi)_{\mathrm{con}} - \alpha \dot\phi \\
     (\delta\dot\phi)_{\mathrm{syn}} &=&  (\delta\dot\phi)_{\mathrm{con}}
     - \alpha \ddot\phi ,
\label{eqn:gaugetrans}
\end{eqnarray}
after noting that $\alpha \simeq (2/3)\Psi/{\cal H}$, during matter domination.
 
We derive the initial conditions in the conformal-Newtonian/longitudinal gauge, for the sake of completeness, which can then be used to evolve Eq.~(\ref{eqn:deltaphitwo}). To obtain the initial conditions in this gauge, we use Eq.~(\ref{eqn:gaugetrans}), where $\alpha=(1/2) \Psi/{\cal H}$ during radiation domination. At early times, deep in the radiation era, we can set the fractional energy-density perturbation in the radiation field to $\delta_r = -2 \Psi$ \cite{Ma:1995ey}, and assume that the equation-of-state parameter $w_{\phi}\to-1$, and changes slowly with time. Furthermore, the pressure and energy density of the scalar are given as
\begin{equation}
\begin{gathered}
p_{\phi}=\frac{1}{2a^2}\dot\phi^2 - V(\phi),\\
\rho_{\phi}=\frac{1}{2a^2}\dot\phi^2 + V(\phi),\\
p_{\phi}+\rho_{\phi}=\frac{1}{a^2}\dot\phi^2,
\end{gathered}
\label{phi_p_rho}
\end{equation}
and the perturbation in the energy density is
\begin{equation}
(\delta\rho_{\phi})_{\text{con}} = \frac{1}{a^2}\dot\phi\dot{\delta\phi} + V'(\phi)\delta\phi - \frac{1}{a^2}\dot\phi^2\Phi,
\end{equation}
while the adiabatic initial conditions require that the entropy density perturbation vanishes at early times, so that
\begin{equation}
S \equiv \frac{\delta\rho_\phi}{\rho_\phi + p_\phi} -  \frac{\delta\rho_r}{\rho_r +
p_r}.
\end{equation}
Combining these assumptions into the gauge-transformation equations, along with the observation that $\delta \phi$ vanishes in the synchronous gauge, we get the initial conditions for the scalar-field perturbations in the conformal-Newtonian/longitudinal gauge,
\begin{eqnarray}
     (\delta\phi)_{\mathrm{con}}  &=& \frac{1}{2}\frac{\dot\phi}{\cal H}\Psi,\\
     (\delta\dot\phi)_{\mathrm{con}} &=&  \dot\phi \Phi -\frac{3}{2}\dot\phi \Psi
     - \frac{1}{2} \frac{a^2 V'}{\cal H} \Psi.
\end{eqnarray}
These initial conditions can also be derived by requiring that $S$ and $\dot{S}$ vanish at early times.


\section{Full expression for the variance of $\alpha T$
cross-correlation}
\label{sec:variance_alphaT_full_derivation} 

If we do not assume $\widehat \alpha_{LM}$ is a Gaussian,
then the full expression for the variance of its
cross-correlation with the CMB temperature becomes a 6-point
correlation function. After applying Wick's theorem and taking
into account the properties of the Wigner 3j symbols to simplify
the terms, the full expression becomes
\begin{equation}
\begin{gathered}
	(\Delta\widehat C^{\alpha T}_L)^{2} =
        \frac{(C_L^{\alpha\alpha,\mathrm{noise}})^{2}C^{BB\text{,
        noise}}_L}{4\pi W^{2}_L} [2(V^L_{LL})^2(C^{TE\text{,
        map}}_L)^2\\
         +
         \sum_{l}[2\frac{(2l+1)^2}{(2L+1)^2}(V^L_{ll})^2C^{EE\text{,
         map}}_lC^{TT\text{, map}}_L\\ 
         + (1+\delta_{lL})(V^L_{lL})^2(C^{TE\text{, map}}_L)^2]\\
         +\sum_{ll'}(1+\delta_{ll'})
         \frac{(2l+1)(2l'+1)}{(2L+1)}(V^L_{ll'})^2C^{TT\text{,
         map}}_LC^{EE\text{, map}}_{l'}].
\end{gathered}
\label{sigma_alphaT_full_expression}
\end{equation}

 \vfill


\end{document}